\documentclass[showpacs,preprintnumbers,amsmath,amssymb]{revtex4}

\usepackage{graphicx}
\usepackage{dcolumn}
\usepackage{bm}
\usepackage{mathrsfs}
\usepackage{amsmath}

\begin{document}


\title{Reply to ``Comment on `Some exact quasinormal frequencies of a massless scalar
    field in Schwarzschild spacetime' ''}

\author{D. Batic}
\email{davide.batic@ku.ac.ae}
\affiliation{%
Department of Mathematics,\\  Khalifa University of Science and Technology,\\ Main Campus, Abu Dhabi,\\ United Arab Emirates}
\author{M. Nowakowski}
\email{mnowakos@uniandes.edu.co}
\affiliation{
Departamento de Fisica,\\ Universidad de los Andes, Cra.1E
No.18A-10, Bogota, Colombia
}
\author{K. Redway}
\email{mnowakos@uniandes.edu.co}
\affiliation{
Departament of Physics,\\ University of the West Indies,\\ 
Mona Campus, Kingston\\
Jamaica
}%

\date{\today}

\begin{abstract}
We explain why the analysis in our paper [Phys. Rev. D{\bf{98}}, 024017 (2018)] is relevant and correct.
\end{abstract}

\maketitle
The comment on our paper \cite{DB} is essentially based on one major claim, namely that our conclusion about the existence of a new branch of quasinormal modes arises from a misleading notation. This claim is easy to dispose of as erroneous. Let us impose that the coefficient $\beta_n$ in the recurrence relation (17) in \cite{DB} vanishes for some choice of a nonnegative integer $j$. Then, we find
\begin{equation}\label{cpm}
c_{(\pm),j\ell}=-j+\frac{1}{2}(\pm)\frac{i}{2}\sqrt{1+2\lambda},\quad\lambda=\ell(\ell+1).
\end{equation}
In order to keep a clear notation, $(\pm)$ refers to the roots of the equation $\beta_j=0$ while $\pm$ is used to distinguish among the particular solutions of the recurrence relation (17) in \cite{DB}. Proceeding as in \cite{DB}, it is possible to find a spectral family of frequencies  
\begin{equation}\label{qnmsII}
\omega_{(\pm),j\ell}=(\pm)\frac{\kappa}{4}\sqrt{1+2\lambda}-i\frac{\kappa}{2}\left(j+\frac{1}{2}\right),\quad j=0,1,\cdots,
\end{equation} 
where for instance, $\omega_{(+),j\ell}$ is obtained from $c_{(-),j\ell}$ using (14) in \cite{DB}. Both branches ensures that the wave function has an exponential decay in the time variable and its radial part diverges exponentially at the event horizon and asymptotically far away (see (3) and (7) in \cite{DB}). If we substitute (\ref{cpm}) into the recurrence relation (17) in \cite{DB}, the Birkhoff-Adams asymptotic theory for second order linear difference equations ensures that our recurrence relation admits two linearly independent solutions with asymptotic representations
\begin{eqnarray}
a_{\pm(\pm)n,j\ell}&=&e^{\pm\gamma_{(\pm)j\ell}\sqrt{n}}n^{\alpha_{(\pm)j\ell}}\sum_{s=0}^\infty\frac{\mathfrak{c}_{\pm(\pm)s,j\ell}}{n^{s/2}},\quad
\alpha_{(\pm)j\ell}=-\left(1+\frac{j}{2}\right)(\pm)\frac{i}{4}\sqrt{1+2\lambda},\label{aeaf}\\
\gamma_{(\pm)j\ell}&=&L_{j\ell}\left[(\pm)\sin{\frac{\delta_{j\ell}}{2}}+i\cos{\frac{\delta_{j\ell}}{2}}\right],\quad
L_{j\ell}=2\sqrt[4]{1+(2j+1)^2+2\lambda},\quad \delta_{j\ell}=\arctan{\frac{\sqrt{1+2\lambda}}{1+2j}}\in(0,\pi/2),\label{L}\\
\mathfrak{c}_{\pm(\pm)0,j\ell}&=&1,\quad
\mathfrak{c}_{\pm(\pm)1,j\ell}=\pm\frac{16c_{(\pm),j\ell}^2-8c_{(\pm),j\ell}-17-48\lambda}{48\sqrt{c_{(\pm),j\ell}-1}},
\end{eqnarray}
where we used (14) in \cite{DB} to further simplify the expression for $\mathfrak{c}_{\pm(\pm)1,j\ell}$. By means of the asymptotic expansion (\ref{aeaf}) we find that
\begin{equation}\label{ratio}
\left|\frac{a_{\pm(\pm)n+1,j\ell}}{a_{\pm(\pm)n,j\ell}}\right|=1\pm(\pm)\frac{L_{j\ell}}{2\sqrt{n}}+\mathcal{O}\left(\frac{1}{n}\right)
\end{equation}
with $L_{j\ell}$ defined as in (\ref{L}). At this point a couple of remarks are in order. First of all, the combinations $+(+)$ and $-(-)$ in the above expression must be disregarded because they lead to divergence at space like infinity. Convergence in the cases $+(-)$ and $-(+)$ in (\ref{ratio}) is ensured by the generalized Gauss criterion proved in \cite{DB}. In the case we consider $c_{(-),j\ell}$, the general solution of the recurrence relation reads
\begin{equation}\label{ff}
a_{(-)n,j\ell}=\rho_1 a_{+(-)n,j\ell}+\rho_2 a_{-(-)n,j\ell}.
\end{equation}
with $\rho_1$ and $\rho_2$ arbitrary constants. According to (\ref{aeaf}) the particular solution $a_{-(-)n,j\ell}$ will blow up as $n\to\infty$. However, the initial condition $a_0=1$ gives the freedom of demanding that $a_{+(-)0,j\ell}=1$ and $a_{-(-)0,j\ell}=0$. Hence, we can choose $\rho_2=0$, and we are left only with the convergent particular solution. A similar reasoning applies in the case we work with $c_{(+),j\ell}$. At this point, one can proceed as in (\cite{DB}) and verify that after substitution of $c_{(\pm),j\ell}$ into (16) and (17) in \cite{DB} the initial conditions are satisfied and the recurrence relation itself does not give rise to an under/overdetermined system of equations for the expansion coefficients. The above treatment shows that the formula (\ref{qnmsII}) indeed satisfies conditions I., II., and III. introduced in \cite{DB} to define a quasinormal mode. This refined analysis actually shows that there is one more branch of quasinormal frequencies that went missed in \cite{DB}.\\
A dubious aspect of the numerical analysis performed by the author is that after having rewritten the massless scalar wave equation using hyperboloidal coordinates (see equation (4) in the comment), the corresponding solution $V(\tau,\sigma)$ is first expressed in terms of a guess given by the sum of an infinite series with variable coefficients plus an improper integral (see equation (123) in \cite{Ans}), and then, it is studied numerically. The disturbing feature of this approach is that the aforementioned Ansatz is defined by the authors in \cite{Ans} as a conjecture and they underline that a strict mathematical proof of it remains a challenging task which is far from the scope of their work. For this reason the authors verify numerically its validity for some cases. In our opinion, using a result derived from a conjecture to disprove a result obtained from a solid mathematical proof cannot invalidate the latter, and therefore, the claim raised by the author should not be taken seriously.\\ 
Regarding the numerical counter-example offered by the author let us consider the same numerical example considered in his comment. We fix $j=0$ and $\ell=2$, and we use (37) in \cite{DB} to obtain
\begin{equation}\label{c}
c_{(-),02}=\frac{1}{2}-i\frac{\sqrt{13}}{2}.
\end{equation}
The explanation provided after equation (\ref{ff}) in this reply clearly shows that for the numerical values $c_{(-),j\ell}$ one should work with the solution $a_{+(-)n,j\ell}$ of the recurrence relation, more precisely
\begin{equation}\label{m}
a_{+(-)n,02}=e^{-\gamma_{(+)02}\sqrt{n}}n^{\alpha_{(+)02}}\sum_{s=0}^\infty\frac{\mathfrak{c}_{+(-)s,02}}{n^{s/2}},\quad\gamma\approx 1.6557-i 2.1775.
\end{equation}
It is gratifying to observe that $a_{+(-)n,02}$ exhibits asymptotically an exponential decay, and that the same conclusion is obtained for other values of $j$ and $\ell$ satisfying the constraint (48) in \cite{DB}. Hence, by means of the same numerical example suggested by the author we can conclude that his claim that (\ref{qnmsII}) will give rise to an exponential growth is not correct. Furthermore, it is also straightforward to verify that the wave function will diverge at the horizon and asymptotically at infinity. This can be also seen by means of the numerical example above. By means of (12) in \cite{DB} we find as expected that the wave function diverges at the event horizon and asymptotically at infinity, as $\psi_{\omega_{(+),02} 2}(x)\approx (x-1)^{-0.25-i0.901}$ and $\psi_{\omega_{(+),02} 2}(x)\approx x^{0.25+i0.901}e^{(0.25+i0.901)x}$, respectively. In the last numerical counter-example the author of the comment claims that the following numerical values  
\begin{equation}\label{dd}
c_{(\pm)}=1+x\left(-\frac{1}{2}\pm\frac{\sqrt{3}}{2}i\right),\quad x\in\mathbb{R},\quad x>1
\end{equation}
`meet all the requirement (including the inequality). They are clearly not another new branch of QNMs because the sequences satisfying conditions (I) and (III) are linearly independent.' The author seems to forget a key aspect in the derivation of our result, namely that the branch of quasinormal frequencies has been obtained by imposing that the coefficient $\beta_n$ in the recurrence relation vanishes for some choice of a non-negative integer. A trivial substitution of (\ref{dd}) into the expression for $\beta_n$ shows that the latter will never vanish (at least as long as $x$ is real). More precisely, we have
\begin{equation}
\beta_n=(1-i\sqrt{3})x^2+(1+i\sqrt{3})(2n+1)x-n^2-(n+1)^2-\ell(\ell+1).
\end{equation}
The above consideration shows that the numerical counter-example provided by the author is pointless. We would like to stress the fact that all conclusions obtained in the comment and in \cite{Ans} are of numerical nature and they are based on a sequence of weak and questionable arguments. To conclude, we would like to mention the work of \cite{Mash} where it has been also shown by a different approach that Leaver's continued fraction \cite{Leaver} does not encode all possible quasinormal modes. In that case, a new family of quasinormal frequencies was obtained by observing that Leaver's continued fraction breaks down whenever the infinite power series used to represent the radial wave function reduces to a polynomial. All the reasoning of the comment under consideration, especially the implicit claim that a quasinormal mode should be discarded if not found numerically, can be now applied to the results of reference \cite{Mash}. Such a claim would lead to strange contradictions. In general, over several decades many methods have been engineered (see references in \cite{DB}) to find approximate numerical values of the quasinormal modes. But the special quasinormal modes found in \cite{DB} and \cite{Mash} depend on the exact form of the black-hole potential for massless scalar and gravitational perturbations, respectively. Therefore, we will not expect they appear as a quasinormal mode in any method that replaces the black-hole potential by an approximate one or it makes use of unproved conjectures as in \cite{Ans}. Moreover, any numerical algorithm starts from a set of equations which, instead of being solved analytically, are treated numerically. This input constraints the generality of any method. In particular, if, as mentioned above, Leaver's method does not exhaust all possible quasinormal modes, any algorithm based on this method must necessarily miss some branch of quasinormal frequencies. In view of the evidence given above, we conclude that we  did  not  misinterpret  anything  in  connection  with  our  result. We presented a branch of quasinormal modes and proved that it satisfies all necessary conditions.


\begin{thebibliography}{999}
\bibitem{DB}
D. Batic, M. Nowakowski and K. Redway, {\it{Some exact quasinormal frequencies of a massless scalar field
in Schwarzschild spacetime}}, Phys. Rev. D{\bf{98}}, 024017 (2018) 
\bibitem{Ans}
M. Ansorg and R. Panosso Macedo, {\it{Spectral decomposition of black-hole perturbations on hyperboloidal slices}}, Phys.Rev. D{\bf{93}}, 124016 (2016)
\bibitem{Mash}
H. Liuy and B. Mashhoon, {\it{On the spectrum of oscillations of a Schwarzschild black hole}}, Class. Quantum Grav. {\bf{13}}, 233 (1996)
\bibitem{Leaver}
E. W. Leaver, {\it{An Analytic Representation for the Quasi-Normal Modes of Kerr Black Holes}}, Proc.  R. Soc. London A{\bf{402}}, 285 (1985)
\end{thebibliography}
\end{document}